# Magnetic dipolar interaction between correlated triplets created by singlet fission in tetracene crystals

Rui Wang,[1,2] Chunfeng Zhang,[1,2,*] Bo Zhang,[1,2] Yunlong Liu,[1,2]

Xiaoyong Wang,[1,3] and Min Xiao[1,2,4*]

[1]National Laboratory of Solid State Microstructures, School of Physics, Nanjing University, Nanjing 210093, China

[2]Synergetic Innovation Center in Quantum Information and Quantum Physics, University of Science and Technology of China, Hefei, Anhui 230026, China

[3]Collaborative Innovation Center of Advanced Microstructures, Nanjing University, Nanjing 210093, China

[4]Department of Physics, University of Arkansas, Fayetteville, Arkansas 72701, USA

**Correspondence to:** cfzhang@nju.edu.cn; mxiao@uark.edu

**Singlet fission (SF) can potentially break the Shockley-Queisser efficiency limit in single-junction solar cells by splitting one photo-excited singlet exciton ($S_1$) into two triplets ($2T_1$) in organic semiconductors[1-6]. A dark multi-exciton (ME) state has been proposed as the intermediate connecting $S_1$ to $2T_1$[7-10]. However, the exact nature of this ME state, especially how the doubly-excited triplets interact, remains elusive. Here, we report a quantitative study on the magnetic dipolar interaction between SF-induced correlated triplets in tetracene crystals by monitoring quantum beats relevant to the ME sublevels at room temperature. The resonances of ME sublevels approached by tuning an external magnetic field**

**are observed to be avoided, which agrees well with the theoretical predictions considering a magnetic dipolar interaction of ~ 0.008 GHz. Our work paves a way to quantify the magnetic dipolar interaction in organic materials and marks an important step towards understanding the underlying physics of the ME state.**

The spin-allowed SF process is highly efficient in some organic semiconductors that can be implanted in multiple device architectures to improve the efficiency of solar conversion.[4, 11-13] However, the intrinsic mechanism responsible for the fast fission process is still under intense debate, despite remarkable progresses have recently been made[3, 5-10, 14-20]. The essential role played by the intermediate ME state ($^1(TT)$) has been realized[3, 7-10, 14, 16, 21] with the SF process being described as[21],

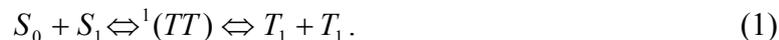

$$S_0 + S_1 \Leftrightarrow {}^1(TT) \Leftrightarrow T_1 + T_1. \tag{1}$$

The doubly-excited ME state, also referred to as the correlated triplet pair, is created with spin coherence when a singlet-excited molecule shares its energy with a neighboring molecule at its ground state. The magnetic interaction between the correlated triplets is insightful for revealing the nature of the ME state by providing valuable information about their spatial separation and the effect of spin coherence[9, 16, 22]. However, it is challenging to quantify the magnetic dipolar interaction because its strength is generally much weaker than the electrostatic counterpart at microscopic scale[23].

Magnetic dipolar interaction is pivotal for many technically-significant processes in organic materials including exciton fission/fusion[21], organic magneto-resistance[24], and organic photovoltaics[25]. In a SF sensitizer, a conceivable approach to quantify the magnetic dipolar interaction is to investigate the interaction-induced energy shift of the ME state. In crystalline tetracene, the energy differences between the ME sublevels can be monitored through quantum beating signals in the singlet population[9, 16, 26]. As depicted in Fig. 1a, exciton fusion from two ME sublevels induces quantum beats in population of the $S_1$ state, manifesting themselves as an oscillation in the time-resolved fluorescence (TRFL) spectrum. After first introduced in 1980s [26], this quantum beat phenomenon in tetracene has recently been re-examined and comprehensively explained by Bardeen's group [9, 16, 27]. Being a direct evidence of the ME state, the quantum beating signal has been regarded as a fingerprint of SF in tetracene [9, 14, 16]. However, the signature of magnetic dipolar interaction has never been captured in tetracene [9, 26].

Here, we exploit a scenario of the interaction-induced anti-crossing to quantify the interaction strength in crystalline tetracene. In our experiments, we manipulate the ME sublevels by applying an external magnetic field to approach the level-crossing resonance [Fig. 1b]. The magnetic dipolar interaction between the correlated triplets causes an avoided level crossing, resulting in two mixed states with an energy gap [Fig. 1c]. The interaction strength is proportional to the gap size that can be precisely evaluated by measuring the frequencies of quantum beats relevant to the resultant upper and lower levels with TRFL spectroscopy [Fig. 1c].

We describe the sublevels of ME state with a spin-dependent Hamiltonian of two correlated triplets. The Hamiltonian includes two isolated triplets (e.g., α and β) and their mutual interaction ($H_{int}$)[9], i.e.,

$$H_{sp}^{TP} = H_{sp}^{\alpha} + H_{sp}^{\beta} + H_{int}. \qquad (2)$$

The Hamiltonian for the magnetic dipolar interaction between the correlated triplet dipoles can be written as[24]

$$H_{int} = X[S^{\alpha} \cdot S^{\beta} - 3(S^{\alpha} \cdot \vec{R}/R_0)(S^{\beta} \cdot \vec{R}/R_0)], \qquad (3)$$

where $S^{\alpha}$ and $S^{\beta}$ are the spin operators of the two triplets, $\vec{R}$ is the displacement vector between the two triplets with the magnitude of $R_0$, and $X$ is the parameter of interaction strength. As described in Supplementary Information (SI), we calculate the energy diagram of the ME sublevels by solving the Hamiltonian of Eq. (2) with a basis of nine eigenstates ($|\phi_{TP}^i\rangle$). In spite of weak magnitude, the interaction part can cause substantial difference from the energy levels predicted by interaction-free model [Fig. S3, SI]. Those sublevels with non-zero mapping to the $S_1$ state ($|\langle S_1|\phi_{TP}^i\rangle|^2 \neq 0$) involve in the exciton fission/fusion processes and directly contribute to the quantum beats[9].

Theoretical calculation predicts a sublevel crossing with a magnetic field of ~ 420 Gauss applied along the x axis [Fig. S3, SI]. The mutual interaction opens a gap of $\Delta \approx 2X$ at the level-crossing resonance [Fig. S3b & d, SI]. Since the gap size is small, it is difficult to directly measure the beat frequencies in time-domain due to damping of the oscillations[9]. To overcome this obstacle, we slightly tilt the external magnetic field with a small angle ($\theta$) relative to the x axis in the xy plane [Fig. 2a] to introduce an extra perturbation ($\delta E_y$) in the Hamiltonian operator. The gap size then increases to be $\Delta \approx 2X + 2\delta E_y$ [Fig. S6, SI], so the evaluation becomes more accurate because the visibility of quantum beat is significantly enhanced in TRFL trace.

Figure 2b plots the TRFL traces recorded under different magnetic fields at $\theta \sim 2°$. An anomalous oscillation with the frequency of ~ 0.05 GHz emerges due to the newly-opened gap when the field is close to 420 G [Inset, Fig. 2b]. The field-dependent amplitudes and frequencies of quantum beats [Fig. 2c] are compared with theoretical calculations [Fig. S5, SI] to further confirm the above assignments. The interaction-free model well explains the observed high-frequency beats (> 0.1 GHz) [Fig. S5], but fails to account for the anomalous part (~ 420 Gauss) without considering the opened energy gap ($\Delta$). A control experiment has been performed with the magnetic field applied at z direction [Fig. S9, SI]. In this configuration, the anomalous beat no longer exists since no level crossing is found [Fig. S8, SI], verifying that the low-frequency quantum beat is indeed a result of gap opened by the interaction and perturbation.

Next, we analyze the θ-dependent oscillatory components to quantify the interaction strength [Fig. 3a]. We assume the displacement vector ($\vec{R}/R_0$) to be along the nearest neighbor orientation[8, 15]. The resonant energies of all nine ME sublevels are calculated by exactly diagonalizing the Hamiltonian (2) and indexed by their zero-field energies. When the magnetic dipolar interaction is included [Fig. S3, SI], the avoided crossing occurs between the 3 & 4 sublevels [Fig. 3c] as well as the 6, 7

& 9 sublevels [Fig. 3d] at ~ 420 Gauss. The mapping products of $\left|\left\langle S_1 | \phi_{TP}^i \right\rangle\right|^2$ for selected sublevel pairs, which directly reflect the beating amplitudes, are compared in Fig. 3e & 3f, suggesting the major contribution from the 3 & 4 sublevels. In Fig. 3b, the measured peak oscillation frequency of anomalous beating signal as a function of the tilt angle θ is compared to the theoretical curves of energy separation between 3 & 4 sublevels calculated with different interaction strengths. The perturbation induced by field tilting corresponds to the zero-interaction curve (X = 0) in Fig. 3b with $\delta E_y \propto \theta$ [Fig. S6d, SI]. The disparity of experimental data from the linear dependence on the tilt angle near $\theta \approx 0$ is an evidence of existing magnetic dipolar interaction, which can be best reproduced by calculation with the interaction strength of X = 0.008 GHz [Fig. 3b].

Furthermore, we have also examined the quantum beats at the strong-magnetic-field limit. A field of ~ 3000 Gauss is applied in the xz plane with an angle of Φ to the x axis as shown in Fig. 4a. In this regime, multiple beat frequencies collapse into one frequency, because only two sublevels dominate the mappings to the $S_1$ state [Fig. S4b, SI][27]. The energy alignments of the two sublevels calculated using models with and without interactions are shown in Fig. 4c & 4b, respectively. The two sublevels approach degeneracy at Φ ≈ 69° [Fig. 4b], which is avoided due to the magnetic dipolar interaction [Fig. 4c]. The oscillatory amplitude is plotted as a function of delay time and the angle deviated from the degeneracy value (ΔΦ) in Fig. 4d. The measured ΔΦ-dependent beat frequencies shown in Fig. 4e agree well with the calculated curve having X = 0.008 GHz. This value of interaction strength is in coincidence with that obtained at 420 Gauss and is at the same order of magnitude as that adopted to interpret the data of optically-detected magnetic resonance in bis(triisopropylsilylethynyl)-tetracene films[22]. It is over two orders of magnitude smaller than the energy difference between the sublevels at zero field, which might explain why the signature of this magnetic dipolar interaction has been absent in previous studies[9, 26].

The orientation of displacement vector ($\vec{R}/R_0$) with respect to the z axis of magnetic tensor may affect the evaluation of the interaction strength. We have considered all four possible configurations for two neighboring molecules in ab plane [Fig. S10, SI]. The interaction strength X [Table S2, SI] that best reproduces the experimental data (Fig. 3f & Fig. 4d) is far below the theoretical value for two nearest neighbor molecules [Table S1, SI], but close to the magnitude of interaction between two magnetic dipoles separated at a distance scale of 2~4 molecules. One plausible explanation for the weak interaction is the delocalization effect of photo-excited singlet excitons[28]. The exciton size has been characterized in the order of tens of molecules at low temperature[28], which might also affect the exciton dynamics in certain extent at room temperature[20]. The pair of triplets resulted from delocalized singlet excitons can distribute over a distance of multiple molecules, which can naturally explain the weak magnetic interaction between the correlated triplets.

Overall, the spins of correlated triplets are weakly coupled to the environment, making our method reliable to quantify the magnetic dipolar interaction at room temperature. The experiments presented here mark an important step towards understanding the nature of ME state for the efficient SF process in tetracene. Including the effect of exciton size to elucidate SF may be enlightening to address the debates between a charge-transfer mechanism and coherent superposition model responsible for the generation of intermediate ME states[9, 10, 15, 18]. The quantum beating behavior in crystalline tetracene due to spin coherence of the correlated triplet pairs is sensitive to the magnitude and direction of the magnetic field. These features deserve more in-depth investigations and can stimulate further adventures in applying SF to many new areas of research such as spintronics and quantum information[23, 29, 30].

**Method**s

Tetracene single crystals with thickness of ~ 1 μm and size up to 5×5 mm$^2$ were prepared by the method of physical vapor deposition. The crystallographic axes were determined by X-ray diffraction and polarization microscopy, which were further employed to determine the magnetic axes with a transform matrix as established by electron spin resonance measurements [Details are available in SI]. A rotatable magnetic coil was used for magnetic-field-dependent experiments. The samples were mounted on a multi-axis platform to realize the desired alignment with respect to the magnetic field. For optical characterization, the second harmonic generation (400 nm) of the pulses emitted from a Ti:Sapphire oscillator (800 nm, Vitara, Coherent Inc.) was chosen as the excitation source. The repetition rate was reduced down to 4 MHz by a pulse picker and the excitation flux was kept at a low level (~ 4 nJ/cm$^2$) to avoid the effect of exciton-exciton annihilation. The emission light was collected by a fiber and routed to a spectrograph. The TRFL spectra at 530 nm were recorded with the technique of time-correlated single-photon counting by an avalanche photodiode having a temporal resolution of better than 50 ps. The multi-exponential decay components were subtracted from TRFL traces prior to Fourier transform to extract the frequencies and amplitudes of quantum beating signals. In theoretical treatments, the resonant energies of ME sublevels are calculated by exact diagonalization of the full spin-dependent Hamiltonian. Quantum beat in the detected TRFL signal is modelled by the dynamics of the density-matrix elements governed by the quantum Liouville equations (see SI for details).

## Acknowledgements

This work is supported by the National Basic Research Program of China (2013CB932903 and 2012CB921801, MOST), the National Science Foundation of China (91233103, 61108001, 11227406 and 11321063), Fundamental Research Funds for the Central Universities and the Priority Academic Program Development of Jiangsu Higher Education Institutions (PAPD). We acknowledge Vitaly Podzorov, Yuanzhen Chen, and Jonathan Burdett for providing valuable information for crystal growth and sample characterization, Shimeng Zhang, and Xiaoling Zhai for single crystal growth, Haifeng Ding for help on the magnetic field setup, Di Wu for stimulated discussion and Xuewei Wu for his technical assistant.

## Author Contributions

C.Z. and M.X. conceived and designed the experiments. R.W., B.Z. and Y.L. performed the experiments. R.W. and C.Z. analyzed the data. R.W., C.Z., X.W. and M.X. co-wrote the manuscript.

## Additional Information

The authors declare no competing financial interests. Correspondence and requests for materials should be addressed to C.Z. or M.X..

# Figures

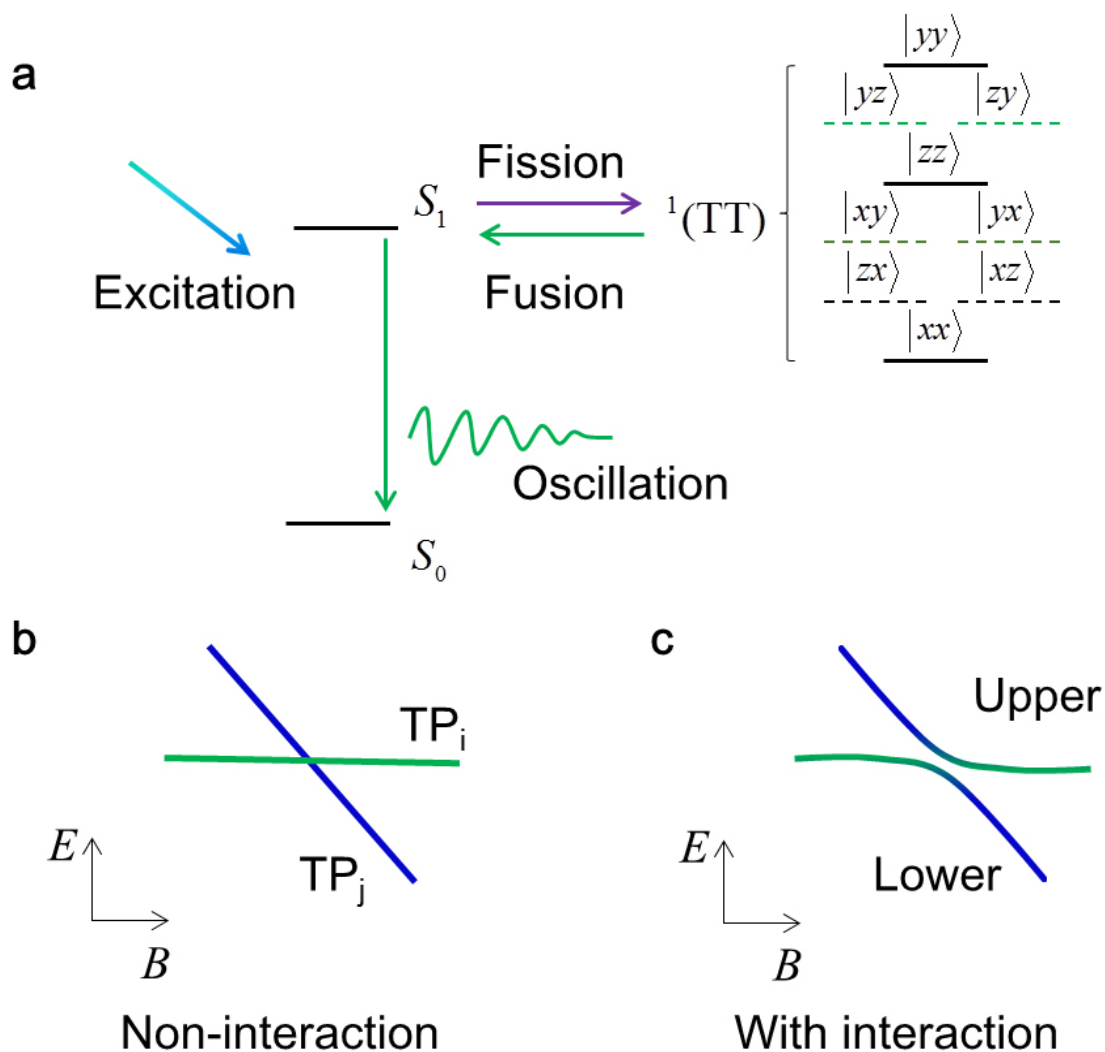

**Figure 1 | Schematic diagram of interaction-induced anti-crossing among ME sublevels. a,** The scenario of the quantum beats related to the ME sublevels in tetracene crystals at zero field. Exciton fusion from sublevels of ME state induces the delayed fluorescence where quantum beats manifest themselves as oscillations in the TRFL spectrum. **b** and **c** plot the field dependences of two near-resonant sublevels for the cases of non-interaction and with interaction between correlated triplets, respectively. With magnetic interaction, the emergence of mixed states results in an avoided crossing with separated upper and lower branches (**c**).

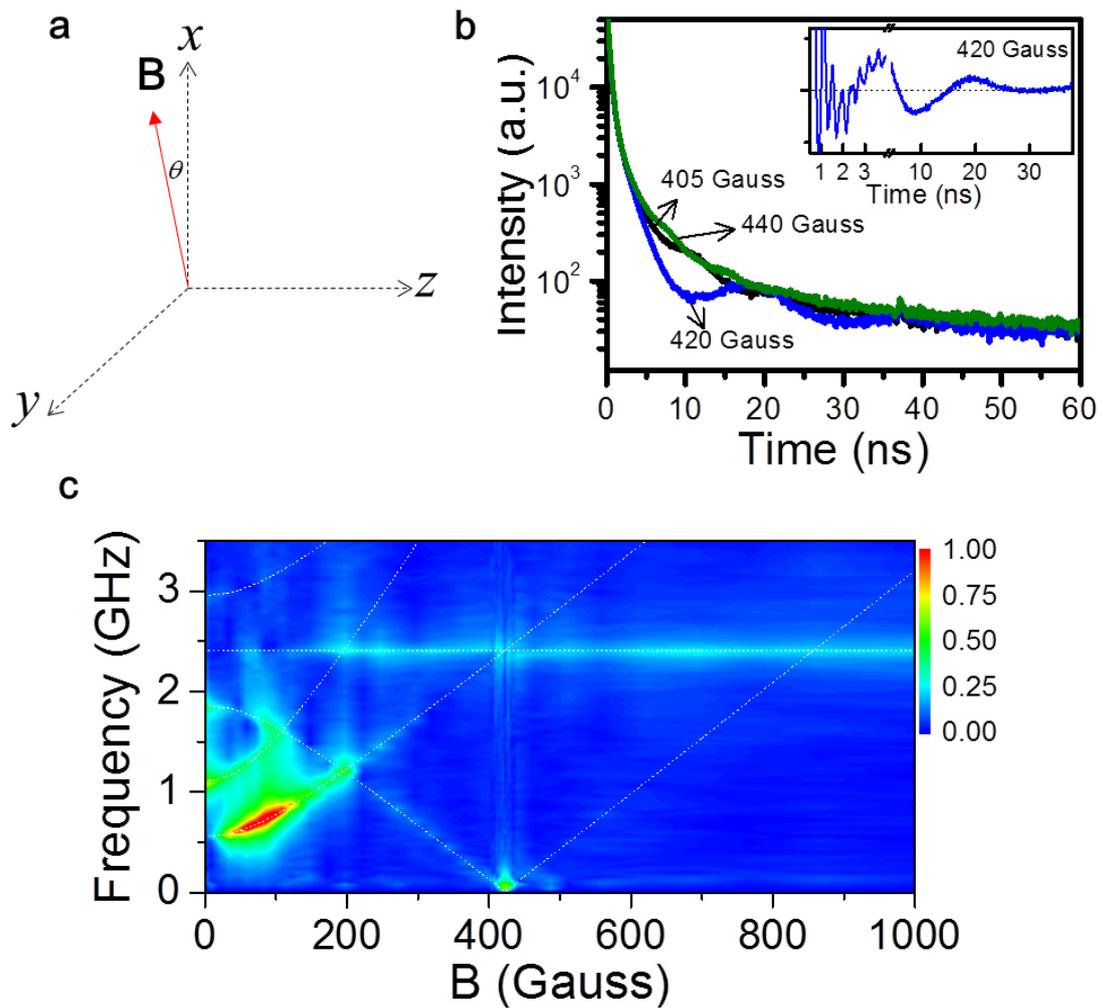

**Figure 2| Magnetic-field-dependent quantum beats. a,** Schematic diagram of the external magnetic field applied at the xy-plane with a small angle ($\theta$) tilted relative to the x-axis. **b,** The decay curves of fluorescence dynamics with three different external magnetic field value tilted at $\theta = 2°$ near the resonance of level crossing. Inset shows the oscillation part obtained by subtracting the multi-exponential decay components from the raw data recorded at 420 G. **c,** The relative amplitudes of quantum beats are plotted as functions of the beat frequency and the field magnitude. The dashed lines indicate the expected beat frequencies calculated with interaction-free model.

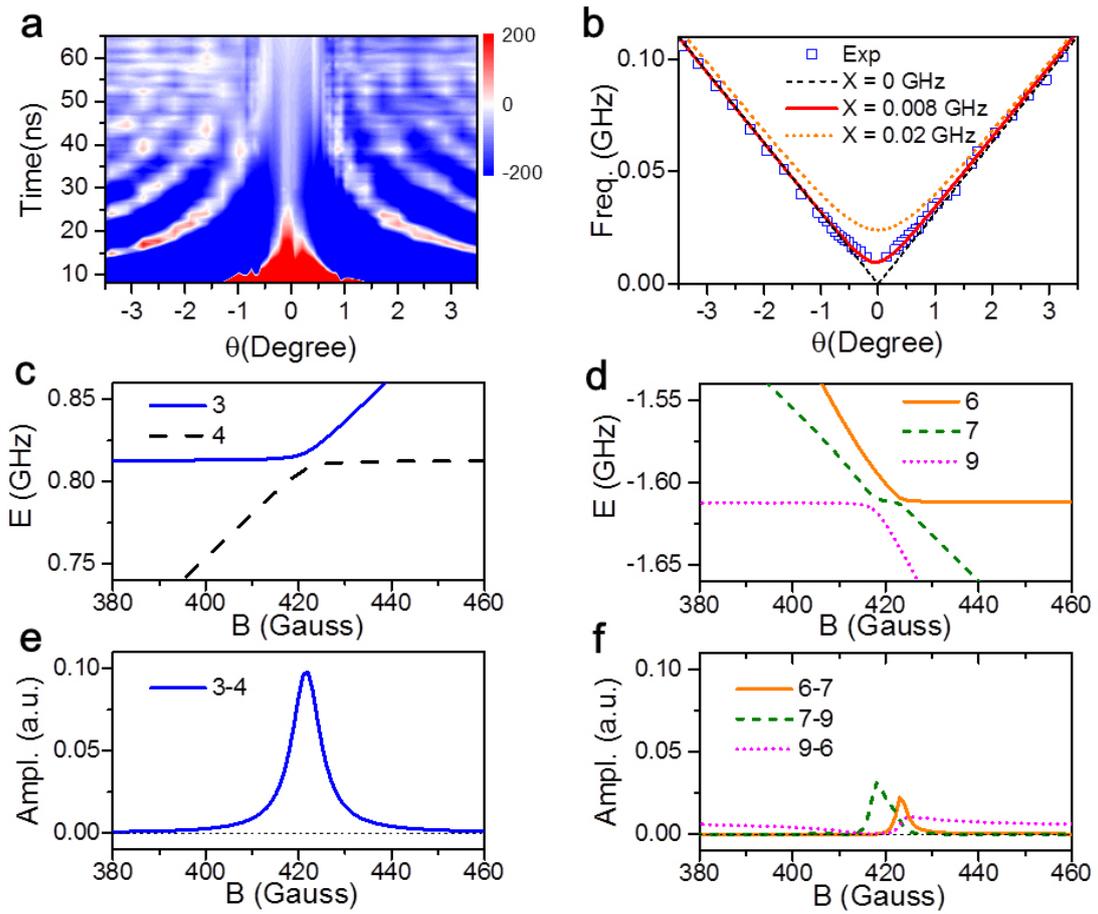

**Figure 3 | Quantum beats near level-crossing resonances. a,** The oscillation components are plotted vs the delay time and tilt angle (θ). The external field is set at 420 Gauss. **b,** Experimental results of θ-dependent beat frequencies are compared with theoretical calculated curves considering different strengths (X) of magnetic dipolar interaction. **c-f,** Theoretical considerations of the interaction-induced anti-crossing of various ME sublevels [For details, see SI]. At the resonance field of ~ 420 Gauss, the anti-crossing occurs in two cases, resulting in the mixed states of 3 & 4 sublevels (**c**) and the mixed states of 6, 7, & 9 sublevels (**d**). The field-dependent amplitudes of low-frequency quantum beats calculated approximately as the mapping products of selected sublevel pairs for these two cases are plotted in **e** and **f**, respectively.

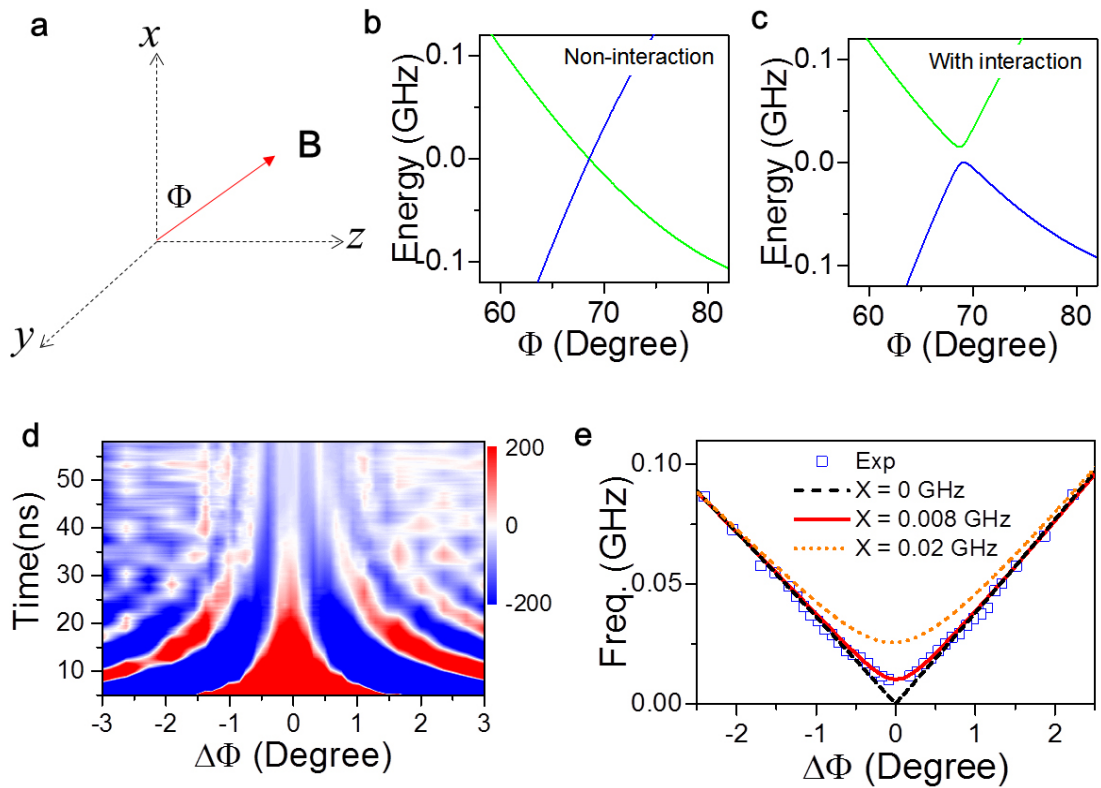

**Figure 4 | Quantum beats at the strong-magnetic-field limit. a,** Schematic diagram of a strong field applied at the xz-plane having an angle Φ with respect to the x-axis. In this strong-field limit, only two dominant sublevels are involved in the fission and fusion processes. **b** and **c** plot the calculated energy alignments of two sublevels for the cases of non-interaction and with interaction between correlated triplets, respectively. The level crossing occurs at Φ ≈ 69°. **d,** The oscillation components are plotted vs the delay time and the angle (ΔΦ) deviated from the resonant value. The magnitude of the external field is ~ 3000 Gauss. **e,** The experimentally measured beating frequencies are plotted as a function of ΔΦ under the strong-field limit. The curves are theoretically calculated results with different strengths of magnetic interaction.